\documentclass[proceedings]{JHEP}
\usepackage{epsfig}
\conference{Heavy Flavours 8, Southampton, UK, 1999}

\renewcommand{\a}{\alpha}
\renewcommand{\b}{\beta}

\newcommand{\la}{\lambda}

\renewcommand{\S}{\Sigma}
\newcommand{\La}{\Lambda}

\newcommand{\bea}{\begin{eqnarray}}
\newcommand{\eea}{\end{eqnarray}}
\newcommand{\beq}{\begin{equation}}
\newcommand{\eeq}{\end{equation}}
\newcommand{\nn}{\nonumber}
\newcommand{\fr}{\frac}
\newcommand{\hl}{\hline}

\title{ A test of HHCPT using magnetic moments of  heavy baryons}

\author{I. Scimemi
\\Dep. de F\'{\i}sica Te\'orica, IFIC, Univ. de Valencia-CSIC, \\
Edificio de Institutos de Investigaci\'on de Paterna\\
Aptdo. de Correos 2085 E-46071 Valencia, Spain 
\thanks{
I  thank A. Della Riccia Foundation (Florence, Italy) for support. 
I acnkowledge  M.C. Ba\~nuls, J. Bernab\'eu, V. Gim\'enez, A.Pich
 for collaboration.
Preprint FTUV/99-76, IFIC/99-79 }
}

\preprint{ FTUV/99-76, IFIC/99-79}
\abstract{
First non-trivial chiral corrections  to the  magnetic moments 
 of  triplet ($T$)  and  sextet ($S^{(*)}$) heavy baryons are  presented
using    Heavy Hadron Chiral Perturbation Theory.
The contributions  are calculated 
 up to order ${\cal O}(1/(m_Q \Lambda_\chi^2))$ 
for $T$-baryons and  up to order  ${\cal O}(1/\Lambda_\chi^2)$
for  $S^{(*)}$-baryons.
The renormalization  of the chiral loops 
 is discussed  and 
 relations among  the magnetic moments of different hadrons  are provided. 
}

\begin{document}

\section{Introduction}
In some kinematical regions, which are not far from the chir\-al 
and hea\-vy quark lim\-its, both
Chir\-al Per\-tur\-ba\-tion Theo\-ry (ChPT) and Heavy Quark Effective Theory (HQET)
 can be combined to\-ge\-ther to con\-struct an ef\-fec\-tive 
Lag\-ran\-gian which
 describes soft interactions of hadrons containing a single heavy quark~\cite{wise}-\cite{cho}.
The  outcoming theory takes  the name  of Heavy Hadron Chiral
 Perturbation Theory (HHCPT).
The basic fields of  HHCPT are heavy hadrons and  light mesons.
In ref.~\cite{chogeo},
 the formalism is extended to include also electromagnetism.
We use this hybrid effective Lagrangian to calculate 
the magnetic moments (MM)
of some baryons containing a $c$ or a $b$ quark.
We find that  within HHCPT,  it is possible to  establish relations among the MM of 
 all this baryons. This measurement  can provide  a useful test for HHCPT. 
In this paper we review the issue as it has been presented in ref.~\cite{nos}.

The light degrees of freedom in the ground  state  of a 
baryon  with one heavy quark can be either in
a $s_l=0$ or in a $s_l=1$ configuration.
The first one cor\-responds  to $J^P=\frac{1}{2}^+$ baryons and   
are annihilated by $T_i(v)$ fields which transform
as a $\bar{\bf3}$ under the chiral $SU(3)_{L+R}$ and as a doublet under 
the HQET $SU(2)_{v}$.
In the second case, $s_l=1$,
the  spin of the heavy quark and the light
 degrees of freedom combine together to form
$J^P=3/2^+$ and $J=1/2^+$ baryons which are degenerate
in  mass  in the $m_Q\rightarrow \infty$ limit.
 The spin-$\frac{3}{2}$ ones are annihilated
by the Rarita-Schwinger field $S_{\mu}^{* ij}(v)$ while the spin-$\frac{1}{2}$
baryons are destroyed by the Dirac field $S^{ ij}(v)$. 

The particle assignement  for the  $J= 1/2$ charmed baryons of 
the $\bar{\bf 3}$ and {\bf 6} representations is
$
(T_1,T_2,T_3)=(\Xi^0_c,-\Xi^+_c,\Lambda^+_c) \ ,
$
\bea
\nonumber
S^{i j} &=& \left ( \begin{array}{ccc} 
\Sigma^{++}_c& \sqrt{\frac{1}{2}} \Sigma^+_c & \sqrt{\frac{1}{2}} \Xi^{+'}_c \\
\sqrt{\frac{1}{2}} \Sigma^{+}_c &\Sigma^0_c & \sqrt{\frac{1}{2}} \Xi^{0'}_c \\
\sqrt{\frac{1}{2}} \Xi^{+'}_c & \sqrt{\frac{1}{2}} \Xi^{0'}_c &
 \Omega^0_c 
\end{array}
\right )\ ,
\eea
and the cor\-res\-pon\-ding $b$-baryons are \\
$
(T_1,T_2,T_3)=(\Xi^-_b,-\Xi^0_b,\Lambda^0_b) \ ,
$
\bea
S^{i j} &=& \left ( \begin{array}{ccc} 
\Sigma^{+}_b& \sqrt{\frac{1}{2}} \Sigma^0_b & \sqrt{\frac{1}{2}} \Xi^{0'}_b \\
\sqrt{\frac{1}{2}} \Sigma^{0}_b &\Sigma^-_b & \sqrt{\frac{1}{2}} \Xi^{-'}_b \\
\sqrt{\frac{1}{2}} \Xi^{0'}_b & \sqrt{\frac{1}{2}} \Xi^{-'}_b &
 \Omega^-_b 
\end{array}
\right ) \ .
\nonumber
\eea
The  $J=3/2$  partners  of these   $S$ baryons   
 have the same $SU(3)_V$ assignement in $S_\mu^{*ij}$.

The  chiral Lagrangian describing the soft \\ 
 had\-ron\-ic and electromagnetic 
interactions of the\-se baryons in the infinite heavy quark mass limit
 can be found in ref.~\cite{nos} and we refer to this paper for its complete
 expression and notation.
Here we  point out only some considerations.

First of all, there are no MM terms in the
lowest order Lagrangian. Therefore, the contributions
to the MM come from:\\
1) the  order ${\cal O} (1/\Lambda_{\chi})$ terms in the baryon chiral Lagrangian
 \cite{nos,chogeo}. 
In the rest of the paper we will take $\La_\chi =4\pi f_\pi\simeq 1.2$ GeV which fixes the normalization of 
 the unknown couplings $c_i$; \\
2) terms of order $1/m_{Q}$ from the heavy quark expansion which
break both spin and flavour symmetries~\cite{nos,chogeo};\\
3) chiral loops of Goldstone bosons coupled to photons, as described
by the lowest order Lagrangian;\\
4) we have not considered
 contributions of order ${\cal O}(1/(m_{Q}\Lambda_{\chi}))$. 
 For the b--baryons, these 
corrections can be safely neglected. For the c--baryons, however, a simple 
estimate shows that their con\-tri\-bu\-tion cannot be larger than, say, 15\%.
 Moreover also 
$O(1/\Lambda^{3}_{\chi})$  contributions  can be neglected because they are NNLO 
chiral corrections.

The renormalization  of the divergent chiral loops which contribute to the MM  requires the introduction of 
higher order operators.
In the case of $S$ baryons  we find that all divergences and scale dependence 
to ${\cal O}(1/\Lambda_\chi^2)$ 
can be absorbed in a redefinition of  only one  ${\cal O}(1/\Lambda_\chi)$
coupling.
 Results are presented in section~\ref{sec:S}.
 The magnetic  moments of the $T$-baryons are analyzed 
in ref.~\cite{sava}. 
However this analysis does not include all meson loops and the needed 
counterterms are not taken into account.
In section~\ref{sec:T} we provide a consistent calculation of the $T$ magnetic moments to order 
 ${\cal O}(1/(m_Q \Lambda_\chi^2))$.
Finally section~\ref{sec:fin} summarizes our conclusions.

\section{Results for  $S$-baryons ($s_l=1$)}
\label{sec:S}

We  define the magnetic moment operator  for a spin 1/2 baryon $B$  and
 a spin 3/2 baryon $B^*_\nu$  respectively as
\beq
-i e \mu(B)
F^{\a\b} \bar{B}\sigma_{\a \b}  B ;\;
-ie \mu(B^*)
 F^{\a\b} \bar{B^*_\mu}\sigma_{\a \b}  B^{*\mu} \ .  
\label{eq:magop}
\eeq
The  leading  contributions from the light- and heavy-quark magnetic interactions are of 
 order ${\cal O}(1/\La_\chi) $ and  ${\cal O}(1/m_Q) $ respectively.
We compute the next-to-leading chiral corrections of order 
${\cal O}(1/\La_\chi^2) $  which originate from the loop diagrams 
shown in fig.\ref{fig:S}.

The resulting MM can be decomposed  as:
\bea
& & \mu (B^{(*)}) =
\fr{1}{72} \left( 6 \fr{Q_{Q}}{m_Q} \mu_{HQE}(B^{(*)})
+\right. \nn \\
& & 
 \fr{16 c_s}{\Lambda_\chi}\mu_{\chi}(B^{(*)}) 
 + 
3 g_2^{2} \fr{\Delta_{ST}}{(4 \pi f_\pi)^2}\mu_{g_2}(B^{(*)})-\nn \\
& & 
\left.
3 g_3^{2} \fr{m_K}{4 \pi f_\pi^2}\mu_{g_3}(B^{(*)}) \right)\ .
\label{eq:magS}
\eea

where $  \mu_i (B)$ and $\mu_i(B^{*})$ are related by 
\beq
\begin{array}{cc}
\displaystyle{\fr{1}{3}}\ \mu_{HQE} (B^{*})=\ \mu_{HQE} (B) =1 & \nn \\
\mu_i (B^{*})=-\displaystyle{\fr{3}{2}}\ \mu_i (B) & i=\chi,\ g_2 ,\ g_3\ .
\end{array}
\eeq

\begin{table}
\begin{center}
\begin{tabular}{|c|c|} \hl
$f_\pi$ & 93 MeV \\
$m_\pi$ & 140 MeV \\
$m_K$ & 496.7 MeV \\
$\Delta_{ST}$ & 168 MeV \\
$m_c$ & 1.3 GeV \\
$m_b$ & 4.8 GeV \\
\hl
\end{tabular}
\caption{Constants used in  numerical estimates.}
\label{tab:cost}
\end{center}
\end{table} 


\begin{table}
\begin{center} 
\begin{tabular}{|c|c|c|}  \hl
  Baryon & 
 $ \mu_\chi  $   &
$\mu_{g_3}   $ 
\\
\hl
$S^{11}$& 2&$1+m_\pi/m_K$    \\
$\sqrt{2} S^{12}$& 1/2&  $1/{2}$  \\
$S^{22}$&$-1$ &$-m_\pi/m_K$    \\
$\sqrt{2}S^{23}$&$-1$&$-(1+m_\pi/ m_K)/2$  \\
$\sqrt{2}S^{13}$& 1/2&${m_\pi}/{(2 m_K)}$   \\
$S^{33}$&$-1$&  $ -1$ \\
\hl
 Baryon &\multicolumn{2}{|c|}{$\mu_{g_2}   $ }\\ \hl
$S^{11}$&\multicolumn{2}{|c|}{$I_\pi$ + $I_K$} \\
$\sqrt{2} S^{12}$ &\multicolumn{2}{|c|}{  $I_K/2$} \\
$S^{22}$ &\multicolumn{2}{|c|}{$-I_\pi$}\\
$\sqrt{2}S^{23}$&\multicolumn{2}{|c|}{ $-(I_\pi+I_K)/2$}\\
$\sqrt{2}S^{13}$&\multicolumn{2}{|c|}{$I_\pi/2$}\\
$S^{33}$&\multicolumn{2}{|c|}{$I_K$}\\
\hl
\end{tabular} 
\caption{Contributions to magnetic moments 
 of spin 1/2 c and b-baryons ($s_l=1$).  }
\label{tab:uno}
\end{center} 
\end{table}

The values of  the $\mu_i(B)$  contributions are reported in  Table~\ref{tab:uno}
 for baryons containing a  $Q$-quark ($Q=c,b$)
where
\bea
I_i&\equiv & I(\Delta_{ST}, m_i) = 2 
\left(-2 +\log{\fr{m_i^{2}}{\mu^{2}}}
\right)+ \nn \\ & &
2 {\sqrt{\Delta_{ST}^{2}- m_i^{2}}\over \Delta_{ST}}
\log{\left(\fr{\Delta_{ST}+ \sqrt{\Delta_{ST}^{2}- m_i^{2}} }{ \Delta_{ST}- 
\sqrt{\Delta_{ST}^{2}- m_i^{2}}
 }\right)} .
\nn
\eea


\FIGURE[pos]{
\epsfig{file=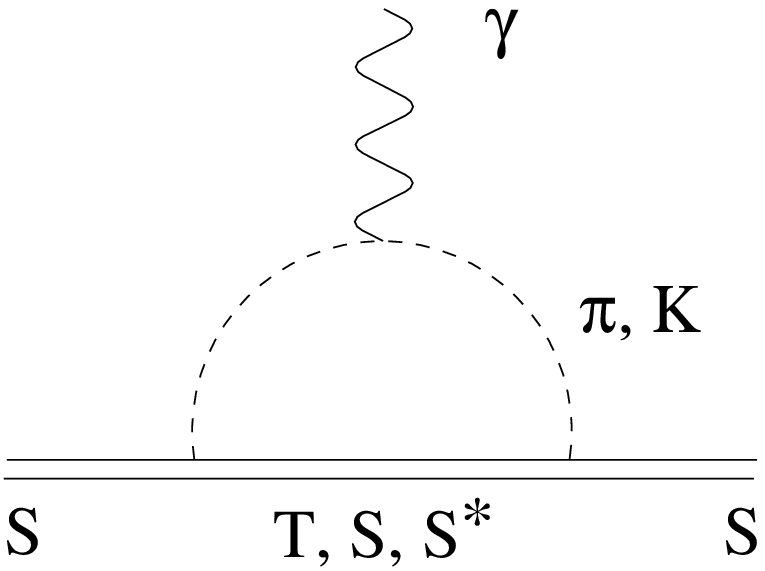}
\caption{Meson loops contributing to $S$-baryons MM.}
\label{fig:S}
             }

We want to stress that due to flavour symmetry, the constants $c_{s}$,
$g_{2}$ and $g_{3}$, and hence the values of  $\mu_{\chi}$, $\mu_{g_{2}}$ and $\mu_{g_{3}}$,
are the same for $c$ and $b$--baryons. The only difference is the contribution 
proportional to $\mu_{HQE}$ due to the different electric charge of the $c$, 
$Q_{c}=+2/3$, and $b$, $Q_{b}=-1/3$, quarks (see Eq.~\ref{eq:magS}).

 The  results  proportional to $g_2^{2}$ are obtained performing a 
one-loop integral (fig.~\ref{fig:S} with an $S$--baryon running in the loop) 
that has to be renormalized.
The divergent part of the integral does not depend on  the pion or kaon mass
and is instead proportional to the  mass of the baryon running in the loop.
If one considers both pion and kaon loops the divergent part  respects
 the $SU(3)$ structure of the  chiral multiplet and  can be canceled
 with an operator of the  form
\beq
\frac{e}{\Lambda_{\chi}^2}
    {\rm tr }\left[\bar{S}_{\mu}\left(v\cdot D S_{\nu}\right) Q
 -\left( v\cdot D\bar{S}_{\mu}\right) S_{\nu} Q
\right] F^{\mu \nu}  \ . 
\label{eq:dseis}
\eeq
This  is the most general dimension--6 chiral and Lorentz 
invariant operator 
constructed out of $S_{\mu}^{i j}$ and $Q\, F_{\mu \nu}$, preserving parity 
and time-reversal invariance
which contributes to MM.
When the equation of motion 
($(v\cdot D)\, S_{\mu}\,=\, \Delta_{ST}\, S_{\mu}$) is applied,
 its contribution
 is of the same form as the term proportional to $c_{s}$.
Thus, the local contribution from the operator in Eq.~\ref{eq:dseis}
 can be taken into account,
through an effective coupling $c_S(\mu)$.
The scale $\mu$ dependence of the loop integrals  is exactly canceled
by the corresponding dependence of the coefficient
$c_S(\mu)$.

The contribution proportional to $g_3^{2}$ in\-volves  a loop int\-egr\-al
 in which a baryon of the $T$-multip\-let is running in the loop.
However,
 as we are in the limit of $m_T\rightarrow \infty$ no mass term
 for these $T$-baryons is present in the lower order  Lagrangian.
This means that the only massive particles running in the loop are 
the light mesons and
 the result of the integral is  convergent and proportional to  
 their mass.

 Using Table~\ref{tab:uno} one  can derive  the following
 linearly  independent relations for  the magnetic moments of spin 1/2
baryons containing a $c$-quark:
\bea
\mu(\S^{++}_c)+\mu(\S^{0}_c)&=& 2\mu(\S^{+}_c)
\nn \\ 
\mu(\S^{++}_c)+\mu(\Omega^{0}_c)&=& 2 \mu(\Xi^{+'}_c)
 \nn \\ 
\mu(\S^{++}_c)+2\mu(\Xi^{0'}_c)&=&\mu(\S^{0}_c)+2\mu(\Xi^{+'}_c)
 \nn \\ 
\mu(\S^{0}_c)+2\mu(\Xi^{+'}_c)&=&\fr{1}{6 m_c}\ . \label{eq:prel}
\eea
Including the spin 3/2 baryons  one can derive six more independent
relations,
\bea
\mu(\S^{++*}_c)+\mu(\S^{0*}_c)&=&2\mu(\S^{+*}_c)  \nn \\ 
\mu(\S^{++*}_c)+\mu(\Omega^{0*}_c)&=&2 \mu(\Xi^{+'*}_c) \nn \\  
\mu(\S^{++*}_c)+2\mu(\Xi^{0'*}_c)&=&\mu(\S^{0*}_c)+2\mu(\Xi^{+'*}_c) \nn \\ 
\mu(\S^{0*}_{_c })+2\mu(\Xi^{0'*}_{_c })&=&3\left(\mu(\S^{0}_c)-2\mu(\Xi^{+'}_c)\right)
 \nn \\
  \fr{2}{3}\mu(\S^{++*}_{c })- \mu(\S^{0}_c)&=&2\mu(\Xi^{+'}_c)-\mu(\S^{++}_c) 
 \nn \\
6\mu(\S^{+}_c)-4\mu(\S^{++}_c) &=& -4\mu(\S^{+*}_c)+\fr{8}{3}\mu(\S^{++*}_c) \ .
\nn \\
\label{eq:prel*}
\eea
The last three equations connect observables  corresponding to
spin 1/2  and  spin 3/2 baryons.

Moreover, it is easy
to deduce  10 analogous equations that  relate
baryons having a $b$-quark
\bea
\mu(\S^{+}_b)+\mu(\S^{-}_b)&=&2\mu(\S^{0}_b)   \nn \\ 
\mu(\S^{+}_b)+\mu(\Omega^{-}_b)&=&2 \mu(\Xi^{0'}_b)  \nn \\
\mu(\S^{+}_b)+2\mu(\Xi^{-'}_b)&=&\mu(\S^{-}_b)+2\mu(\Xi^{0'}_b)  \nn \\ 
\mu(\S^{-}_b)+2\mu(\Xi^{0'}_b)&=&-\fr{1}{12 m_b}  \nn \\
\mu(\S^{+*}_b)+\mu(\S^{-*}_b)&=&2\mu(\S^{0*}_b)  \nn \\ 
\mu(\S^{+*}_b)+\mu(\Omega^{-*}_b)&=&2 \mu(\Xi^{0'*}_b) \nn \\
\mu(\S^{+*}_b)+2\mu(\Xi^{-'*}_b)&=&\mu(\S^{-*}_b)+2\mu(\Xi^{0'*}_b) 
 \nn \\ 
 \mu(\S^{-*}_{_b })+2\mu(\Xi^{-'*}_{_b })&=&3\left(\mu(\S^{-}_b)+2\mu(\Xi^{0'}_b)\right) 
\nn \\
\fr{2}{3}\mu(\S^{+*}_{b })- \mu(\S^{-}_b)&=&2\mu(\Xi^{0'}_b)-\mu(\S^{+}_b)
 \nn\\
6\mu(\S^{0}_b)-4\mu(\S^{+}_b)&=&
-4\mu(\S^{0*}_b)+\fr{8}{3}\mu(\S^{+*}_b) \ ,
\nn \\
& & 
\label{eq:prelb*}
\eea
 and two independent  equations that relate $b$-
 and $c$-baryons
\bea
\mu(\S^{0}_b)-\mu(\S^{+}_b)&=&\mu(\S^{+}_c)-\mu(\S^{++}_c) \nn \\
\mu(\S^{++}_c)-\fr{1}{3}\mu(\S^{++*}_{c })&=& \mu(\S^{+}_b)-\fr{1}{3}\mu(\S^{+*}_{b }) \ .
\nn \\
& &
\eea

From Table~\ref{tab:uno}, we see that the order
 ${\cal O}(1/\Lambda_{\chi})$ and ${\cal O}(1/\Lambda_{\chi}^2)$
 contributions  cancel in the sum of all baryon MM within the sextet.
Therefore,  the average
over the baryon moments  measures the MM of the heavy quark,
$
\langle \mu(S_Q)\rangle =\langle \mu(S_Q^*)\rangle/3={Q_Q}/{12 m_Q}
\ .
$

If one has  got a numerical estimate of the couplings $g_2$ and $g_3$,
  it is possible to derive  a  scale independent relation  between any couple 
of baryons.
The combination 
\beq
\mu(B_1) - \fr{\mu_\chi(B_1)}{\mu_\chi(B_2)}\mu(B_2)
\eeq
is independent of 
 the unknown coupling $c_S(\mu)$  an can then be predicted.
For instance
\bea
& &\mu(\S^{+}_b)+2 \mu(\S^{-}_b)= \fr{1}{24}
\fr{g_3^{2}}{4 \pi f_\pi^{2}} (m_K-m_\pi)- \nn \\
& &
\fr{\Delta_{ST}}{24}\fr{g_2^{2}}{(4 \pi f_\pi)^{2}} 
(I_K-I_\pi)- \fr{1}{12 m_b} \ .
\label{eq:dif}
\eea
The couplings $g_2$ and  $g_3$ 
 have been calculated theoretically. In Table~\ref{tab:gteor}
 we report   the results of these computations.

There exists  an  experimental measurement of $g_3$  from CLEO
 coming from the  decay $\S_c^*\rightarrow \Lambda_c \pi $~\cite{cleo,gnec},
$g_3=\sqrt{3} \,(0.57\pm 0.10)$.
The direct measurement of $g_2$ is not possible at present.
However, the quark model relates its value to $g_3$~\cite{gnec}, yielding
$g_2=1.40 \pm 0.25$.

\begin{table}
\begin{center}
\begin{tabular}{|c|c|c|} \hl
Ref. &
$g_2$ & $g_3$ \\ \hl
~\cite{gural} & 1.88 & 1.53\\
~\cite{gnuc} & $1.5$ & $1.06 $ \\
~\cite{grozi} & $0.83 \pm 0.23$ & $0.67 \pm 0.18$\\
~\cite{zhu} & $1.56\pm 0.3\pm 0.3$ &$0.94 \pm 0.06\pm 0.2$ \\ \hl
\end{tabular}
\caption{Theoretical estimates of $g_2$ and $g_3$.}
\label{tab:gteor}
\end{center}
\end{table} 

In order to get a numerical estimate of the left-hand side of Eq.~\ref{eq:dif}
we set $g_2=1.5 \pm 0.3$ and $g_3=0.99 \pm 0.17$ 
and the rest of the constants
 as in Table~\ref{tab:cost}.
We find 
for our best estimate of  Eq.\ref{eq:dif}
\beq
\mu(\S^{+}_b)+2\mu(\S^{-}_b)=0.23 \pm 0.09 \ {\rm GeV}^{-1} \ . 
\eeq

\section{Results for $T$-baryons ($s_l=0$)}
\label{sec:T}

As the light quarks of the $T$-baryons are  in a $s_l=0$ configuration,
  the contributions to the magnetic moments of these hadrons  are $1/m_Q$  suppressed~\cite{cho}. 
The leading term  is of the form $\mu_{HQE}/m_Q$ and  the  first chiral corrections are
 of order  ${\cal O}(1/(m_{Q}\Lambda_{\chi}))$  and  come from~\cite{sava}
\beq
{\cal{L}}^{\prime}_{(long)} = \fr{c_T}{4 m_Q} \fr{e}{\Lambda_\chi}
{\bar T}^i \sigma_{\mu\nu} Q_{ij} T^j F^{\mu\nu} \ .
\eeq
The contributions of  order ${\cal O}(1/(m_{Q}\Lambda_{\chi}^2))$  have different origin:
\begin{enumerate}
\item
there is 
a divergent contribution~\cite{sava} coming 
 through the chiral loops shown in fig. 2, which is proportional to the 
 explicit mass splitting,
$
 \Delta M_Q= 3\fr{\la_{2S}}{m_Q} \ ,
$
  for the  spin 1/2 and spin 3/2  parts of $S$-baryons~\cite{jenk};
 \item besides, one can consider
 a spin--symmetry breaking operator of $O(1/m_{Q})$; 
\bea
{\cal{L}}^{\prime} =
& & \fr{g'}{m_Q}
 \left [\epsilon_{ijk} \bar{T}^i\sigma^{\mu\nu} (\xi_{\mu})_l^j S_{\nu}^{kl}+
\right.
\nn \\
 & &  \left. \epsilon^{ijk} \bar{S}_{kl}^{\mu} \sigma_{\mu\nu} (\xi^{\nu})_j^l T_i\right ] \ ,
\label{eq:nova}
\eea
which gives rise to divergent loop diagrams, as the one in fig. 2, where one
of the vertices is proportional to $g'$.
\item further, there are finite contributions of the same order coming from 
the $SU(3)$-break\-ing operators
 \bea
& & e \fr{\omega_1}{4 m_Q  \Lambda_\chi^2}
 {\bar T}^i \sigma_{\mu\nu} Q_{il} \chi^l_j T^j F^{\mu\nu} +
\nn \\
& &
e \fr{\omega_2}{4 m_Q  \Lambda_\chi^2} Q_Q
 {\bar T}^i \sigma_{\mu\nu} \chi_{ij} T^j F^{\mu\nu} \ ,
\label{eq:nova2}
\eea
where, in the limit of exact isospin symmetry
$
 \chi={\rm Diag}\left(
m_\pi^2,m_\pi^2, 2 m_K^2- m_\pi^2  
 \right) \ .
$
\end{enumerate}
 As in the case of the $S$-baryons, when all Goldstone boson loops are included,
 the scale $\mu$ dependence of the result of fig. 2 
 is canceled by  the   corresponding  dependence  of an effective  $c_T(\mu)$. 
The  interaction term  of Eq.~\ref{eq:nova}
and the finite terms of Eq.~\ref{eq:nova2} 
were first  taken into account in ref.~\cite{nos}.


\EPSFIGURE[pos]{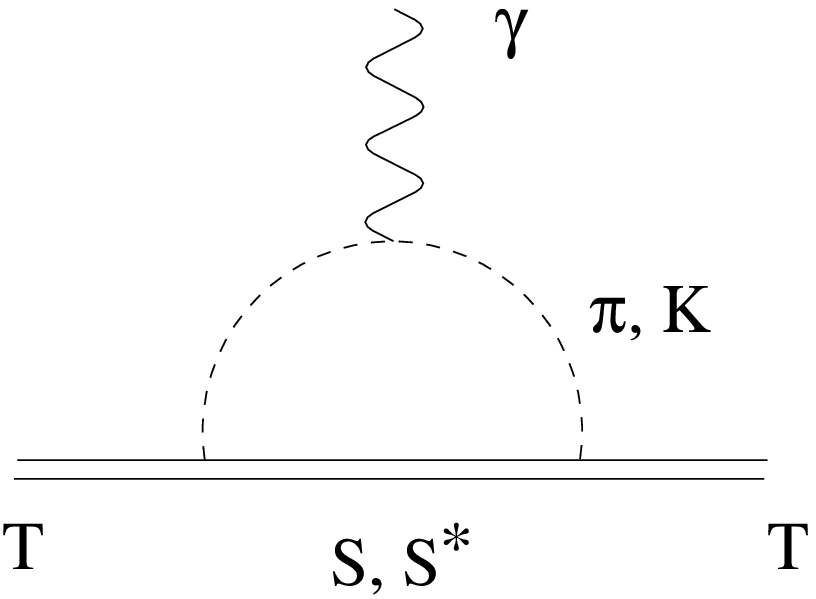}{Meson loops contributing to $T$-baryons MM.}

 Similarly to what we have done  in the previous paragraph we write
 the magnetic moment of $T$-baryons as
\bea
& & \mu (B)= \fr{1}{24 m_Q}\left(-6 Q_{Q}\mu_{HQE}(B)-
 \fr{ c_T}{ \Lambda_\chi }\mu_{T}(B)+ \right. \nn \\
& & g_3^{2} \fr{  3 \la_{2S}}{(4 \pi f_\pi)^2}\mu_{g_3}(B) 
+6 g_3  g' \fr{\Delta_{ST}}{(4 \pi f_\pi)^2}\mu_{g'}(B) + 
\nn \\  
& &
\left.  2 \fr{\omega_1 m_K^2}{\Lambda_\chi^2}
\mu_{\chi_1} (B) -6 Q_{Q}
\fr{\omega_2 m_K^2}{\Lambda_\chi^2 }
\, \mu_{\chi_2} (B)
\right) .
\eea
 The values  of  the  $\mu_i$ are written in Table~\ref{tab:t} where
\beq
J_i=\fr{\partial}{ \partial \Delta_{ST}} 
\left( \Delta_{ST}I(\Delta_{ST}, m_i) \right) \ .
\eeq
Corrections 
to our results for
$T$--baryons  are of order ${\cal O}(1/m_{Q}^{2})$ and hence negligible.

By eliminating the unknown coupling constants,
one  can deduce two independent relations among the magnetic moments of both
T-multip\-lets
\bea
& & m_b\, \mu (\Xi_b^-)-m_c\, \mu (\Xi_c^0) =
m_b\, \mu (\Xi_b^0)-m_c\, \mu (\Xi_c^+) \nn \\
& & m_b\, \mu (\Lambda_b^0)-m_c\, \mu (\Lambda_c^+)- \fr{1}{4} = 
\left( 2\fr{m_K^2}{m_\pi^2}-1\right) \times \nn \\
& & 
\left[ m_b\, \mu (\Xi_b^-)-m_c\, \mu (\Xi_c^0)-\fr{1}{4}\right]
\ .
\eea

In the absence of the $SU(3)$-breaking operators in Eq.~\ref{eq:nova2}, 
the average baryon MM over the $T$ multiplet would be
equal to the heavy quark MM~\cite{sava}.
The result is however corrected by contributions proportional 
to the unknown couplings $\omega_1$ and
 $\omega_2$\footnote{Notice that our definition of MM differs from the one
in ref.~\cite{sava} by a factor $-1/4$.}:
\bea
& & 
\langle \mu(T_Q)\rangle=
\fr{-1}{4 m_c} \left [ Q_Q
+ \fr{2 \omega_1 m_K^2}{9 \La_\chi^2}\left ( 1- \fr{m_\pi^2}{m_K^2}\right )
- \right. \nn \\ & & 
\left . Q_Q \fr{\omega_2 m_K^2}{3 \La_\chi^2}\left ( 2+ \fr{m_\pi^2}{m_K^2}\right )
\right ]  .
\eea

\begin{table}
\begin{center} 
\begin{tabular}{|c|c|c|c|}  \hl
Baryon & 
 $ \mu_T  $   &
$\mu_{g_3}   $ &
$\mu_{g'}   $
\\
\hl \hl
$T_1$ &$4$ &$J_\pi+J_K $ & $I_\pi+I_K$ \\
$-T_2$&$-2$&$-J_\pi$ & $-I_\pi$  \\
$T_3$&$-2$&$-J_K$ &  $-I_K$  \\
\hl 
Baryon &\multicolumn{2}{|c|}{$\mu_{\chi_1}$} &$\mu_{\chi_2}$ \\
\hl 
$T_1$  &\multicolumn{2}{|c|}{$-2m^2_\pi/m_K^2$} & $m^2_\pi/m_K^2$ \\
$-T_2$&\multicolumn{2}{|c|}{$m^2_\pi/m_K^2$ }& $m^2_\pi/m_K^2$ \\
$T_3$&\multicolumn{2}{|c|}{$2-m^2_\pi/m_K^2$} & $2-m^2_\pi/m_K^2$ \\
\hl
\end{tabular} 
\caption{Contributions to magnetic moments 
 of spin 1/2 $T$-baryons ($s_l=0$).  }
\label{tab:t}
\end{center} 
\end{table}


\section{Conclusions}
\label{sec:fin}
The magnetic moments of triplet and sextet hea\-vy baryons have been computed
in the HHCPT.
The calculation    of the $S^{(*)}$-baryons MM  at the order ${\cal O}(1/\Lambda_\chi^2)$
 involves only one new arbitrary constant,
$c_S$.
 Thus it is possible  to derive  relations among the MM of the  hadrons in the same sextet
 where all masses and  effective couplings  are eliminated.
Due to heavy quark symmetry  the MM of the $S$ and  $S^*$ sextets  are also related.
Moreover, as $c$ and $b$  baryons are described by the same arbitrary constants,
we can connect  the MM of the two kinds of hadrons.
The average over  one sextet equals the corresponding heavy quark MM.

In the case of  $T$-baryons the first corrections appear at 
  order ${\cal O}(1/(m_Q \Lambda_\chi^2))$ and  four   arbitrary constants 
are required.
Then  we are left with only two independent relations  which combine
 $c$ and $b$ triplets  and contain $m_c$ and $m_b$.
The average over  one triplet equals the heavy quark MM only in  the absence
of $SU(3)$-breaking operators.

The  measure of the magnetic moments  of heavy baryons represents an
experimental challenge. Nevertheless  several groups are contemplating
the possibility of performing it  in the near future 
(BTeV, SELEX)~\cite{exp}.

\end{document}